\documentclass{article}
\begin{document}

\title{Extraction of $^{137}$Cs by alcohol-water solvents\\
from plants containing cardiac glycosides}
\maketitle

\begin{center}
\linespread{1.3}
S.N. Dzyubak, Yu.I. Gubin \\
{\it \small State Scientific Centre of Medicines, \\
Astronomicheskaya st., 61085 Kharkov, Ukraine}\\
\bigskip
O.P. Dzyubak \footnote{E-mail address: dzyubak@fnal.gov},
P.V. Sorokin, V.F. Popov \\
{\it \small Institute of High Energy Physics and Nuclear Physics, \\
Akademicheskaya st., 61108 Kharkov, Ukraine}\\
\bigskip
A.A. Orlov, V.P. Krasnov\\
{\it \small Ukrainian Research Institute of Forestry and 
Forest Melioration,\\ 
Poushkinskaya st, 61024 Kharkov, Ukraine}\\
\end{center}

\begin{abstract}
As a result of nuclear power plant accidents, large areas receive 
radioactive inputs of $^{137}$Cs. This cesium accumulates in
herbs growing in such territories. The problem is whether the herbs 
contaminated by radiocesium may be used as a raw material for medicine.
The answer depends on the amount of $^{137}$Cs transfered 
from the contaminated raw material to the medicine. 
We have presented new results of the transfer of $^{137}$Cs from 
contaminated  
{\it Digitalis grandiflora Mill.} and  {\it Convallaria majalis L.} 
to medicine. We found that the extraction of $^{137}$Cs depends 
strongly on the hydrophilicity of the solvent.
For example 96.5$\%(vol.)$ ethyl alcohol extracts less 
$^{137}$Cs (11.6 \%) than 40$\%(vol.)$  ethyl alcohol or
 {\it pure water} (66.2 \%). 
The  solubility of the cardiac glycosides is inverse to the solubility 
of cesium, which may be of use in the technological processes 
for manufacturing ecologically pure herbal medicine.\\



\end{abstract}

\section{Introduction}
At the present time, herbal medicines containing cardiac glycosides 
are widely used in the medical treatment of heart disease as 
the equivalent synthetic analogues are not available yet.  
The {\it Convallaria} and {\it Digitalis} species are used
as raw material for the herbal cardiac medicine.\\
As a result of nuclear power plant accidents, large areas receive 
radioactive inputs of $^{137}$Cs. This cesium accumulates in herbs 
growing in such territories. The problem is whether the herbs 
contaminated by radiocesium may be used as a raw material for medicine. 
The answer depends on the amount of $^{137}$Cs transfered 
from the contaminated raw material to the medicine.\\
After the Chernobyl catastrophe a large area, growing many species of 
medicinal plants (such as {\it Convallaria majalis L., Rhamnus catartica L., 
Acorus calamus L., Vaccinium vitisidea L., Vaccinium myrtillus L. etc.}) 
was contaminated with a high concentration of radionuclides. 
The main contaminant (more then 90\% of the overall radioactivity) 
of the Ukrainian Polessie is $^{137}$Cs \cite{Buzun}.
The majority (85 to 97\% ) of $^{137}$Cs is located in the soil layer
at a depth of 0 to 10 cm, where the roots of the medicinal herbs take up
the $^{137}$Cs \cite{Pushkarev}.\\
Various solvents are used in the pharmaceutical industry  to prepare the 
herbal medicinal products. The poperties of these solvents define 
a quantitative and qualitative structure of substances that can be extracted.
The literature gives some often contradictory information 
on the transfer of $^{137}$Cs from medicinal plant raw material 
to water and alcohol. Several studies 
\cite{Orlov96, Krasnov96, Orlov97, Sanarov98, Grischenko90}
have shown that the amount of radionuclide transferred 
from the soil to the raw material varied in wide range 
(from 10 to 250 fold). This difference was mainly determined 
by specific features of the plants 
and depended strongly on the type of soil and climatic conditions 
which occurred during the vegetative period. 
The transfer of $^{137}$Cs from the plant raw material is 24-75\%
for the aqueous medicinal products and is 20-30\% for the alcoholic 
ones \cite{ Antonova89, Prokofiev92, Prokofiev93, Dmitriev91}. 
Grodzinskii et al. \cite{Grodzinskii91} found that 
the specific activity of $^{137}$Cs 
in water extracts was three orders of magnitude less than in 
the initial plant raw material.\\
One can conclude, from the above publications that experiments 
have been mainly devoted to studying the transfer of $^{137}$Cs 
from plant raw material to water and some galenical preparations. 
Systematic studies of the dependency of radionuclide extraction 
efficiency on various types of solvents have not been done yet. 
Therefore the study of radionuclide transfer from soil to herbs and 
from herbs to the herbal medicinal products is very important.

In this paper we present our experimental results on the effect of
solvent type ({\it pure water, 40$\%(vol.)$, 70$\%(vol.)$ and 96.5$\%(vol.)$ 
aqueous ethyl alcohol}) on $^{137}$Cs extraction efficiency.

\section {Methods}

\subsection{ Raw material} 

We studied the medicinal plant species 
(the herbs {\it Digitalis grandiflora Mill}, 
flowers and leaves of {\it Convallaria majalis L.})
containing cardiac glycosides. The raw material was taken from  
an experimental plot of the Povchansk forest area,
Luginy district, Zhitomir region where the soil contamination
by $^{137}$Cs ranged from 296 to 925 $\frac{kBq}{m^2}$
and the radioactivity of the tested plant raw material was 
1.17 -- 50.83 $\frac{kBq}{kg}$ (see Table 1).
All raw plant material was rolled three times in preparation
for the extraction process.

\subsection{ Tinctures and aqueous extracts } 

We used the following solvents: {\it pure water, 
40$\%(vol.)$, 70$\%(vol.)$ and 96.5$\%(vol.)$ aqueous ethyl alcohol} 
without heating and vaporization of solvent. Samples (10.0 $g$ aliquots) 
of raw material were put in glass extractors, filled up 
with the appropriate concentration of solvent and let stand 
for extraction. After 48 hours, the extracts were discharged 
to a flask and the extractors were filled up with a new portion of 
the solvent for further extraction. The operation was repeated 
after 24 hours and 48 hours to obtain the second and third extracts. 
Then the solvent residuals were removed using vacuum and added to 
the extracts. All extracts were combined to obtain
the primary tinctures of 100 $ml$ volume.

\subsection {Decoctions} 

The solvent that we used was hot water. The 10.0 $g$ samples 
of raw material were put in glass vessels, filled up with 
pure water and boiled for 15 minutes. After that these vessels 
were cooled and kept at room temperature for 10 minutes.
The filtered extracts and the solvent residuals were then
combined to give the decoctions of 100 $ml$ volume.

\subsection {Measurement of radioactivity}

The $\gamma$ activity of both the initial raw material 
and the extracts were measured during the experiment. 
The standards of the Health Ministry of Ukraine state that 
the specific activity of $^{137}$Cs in medicinal 
plants must be less than 600 $\frac{Bq}{kg}$ \cite{RD97}.
We needed to measure a small activity and commercial devices 
(like the {\it LP -- 4900B}) have insufficient sensitivity, 
therefore we used the $\gamma$ detector from the Institute for 
Single Crystals (Kharkov, Ukraine). The scintillator ($BGO$) was 
a cylinder with d=40 $mm$ and h=40 $mm$. The detector was calibrated 
with $^{60}Co$ ($E_{\gamma}$=1.173 $MeV$, $E_{\gamma}$=1.332 $MeV$) and 
$^{137}$Cs ($E_{\gamma}$=0.662 $MeV$) sources from 
a standard calibration set before measurements. 
For the $E_{\gamma}$= 0.662 $MeV$  photon the energy resolution 
of the detector was determined to be 13.4\%.
The photopeak efficiency of photon registration 
for the point source was {$\varepsilon $}=0.42. 
The detector and samples were placed inside a 5 $cm$ thick
lead shield to decrease the background. The background at 
the 0.662 $MeV$ peak region was about 0.2 [$\frac{count}{sec}$]. 
The samples were measured directly inside Marinelli's vessel
(100 $ml$ volume). 
The detector efficiency was 31.5 [$\frac{Bq\times sec}{count}$]. 
The activity of $^{137}$Cs in the samples was determined by formula 
A [$\frac{Bq}{kg}$] = $31.5\times N_{ph.p.}$, 
where $N_{ph.p.}$[$\frac{1}{sec\times kg}$] is the number of counts 
in the photopeak of the measured substance.\\
The data acquisition system was assembled in CAMAC standard.
The $\gamma$ spectrum parameters were calculated with the MINUIT program 
from the ROOT package \cite{MINUIT}.\\
All samples were measured over a wide range of photon energy 
(0.1 -- 2.0 $MeV$). Only one peak at 0.662 $MeV$ was detected 
which meant that $^{137}$Cs was present and
therefore we only studied the  transfer of $^{137}$Cs. 
The statistical error of the measurements did not exceed 5\%. 
The absolute error was 15\%  
and depended on the standard calibrated source of $^{137}$Cs.

\section {Results and discussion }

\par

The results of the determination of $^{137}$Cs transfer from the herbs 
to alcohol-water and pure water extracts are presented in Table 1. 
From the table one can see that for the plants investigated, 
the $^{137}$Cs transfer does not depend on the type of raw material 
(within the error limits). The reason for certain scattering of 
the data for $^{137}$Cs transfer for 96.5$\%(vol.)$ alcohol 
can be due to saturation of solvent by cesium.

Studying the dependence of the $^{137}$Cs transfer on the solvent 
hydrophilicity shows that for 96.5$\%(vol.)$ alcohol the transfer 
of $^{137}$Cs to the extract is minimal, ranging from 4.6 to 17.6\%.
For 70$\%(vol.)$ alcohol the transfer reaches 45.3 to 66.5\%. 
In our experiments the maximal amount of $^{137}$Cs  
was extracted by 40$\%(vol.)$ alcohol  (62.8 to 83.2\%) 
and pure water at room temperature (63.0 to 73.0\%).
Decoctions extracted radiocesium similar to 70$\%(vol.)$ alcohol 
(49.3 to 60.8\%) but less than 40$\%(vol.)$ alcohol or pure water 
at room temperature. The reason is that in the process of heating 
some of the cesium chemically interacts 
with the raw material and can not be extracted.

As follows from our results the 96.5$\%(vol.)$ alcohol extracts less 
(about 6 times) $^{137}$Cs than 40$\%(vol.)$. From Table 2 one can see 
that solubility of the cardiac glycosides strongly depends 
on solvent type and the solubility tendency of the cardiac glycosides
is inverse to cesium \cite{Baumgarten}. Methanol extracts the cardiac 
glycosides much more (a factor of 570 for {\it Digitoxin} and 
a factor of 18 for {\it Convallatoxin}) as compared to water.
Thus the ratio of  $^{137}$Cs to the cardiac glycosides extracted 
strongly depends on solvent type.

\section {Conclusion}

We have presented new results on the transfer of $^{137}$Cs from raw 
material to medicine. We have found that the extraction of $^{137}$Cs 
from {\it Digitalis grandiflora Mill.} and {\it Convallaria majalis L.} 
containing cardiac glycosides strongly depends on solvent 
hydrophilicity and where 96.5$\%(vol.)$ alcohol extracts less 
$^{137}$Cs (about 6 times) than 40$\%(vol.)$ or {\it pure water}. 
The  solubility tendency of the cardiac glycosides is inverse to 
that of cesium and this fact can be of use in the technological 
processes for manufacturing ecologically pure herbal medicine.\\

{\bf \Large Acknowledgements}\\

The authors are grateful to S.F. Burachos for use of
the $BGO$ scintillator.

\newpage

Table 1 \\
$^{137}$Cs transfer from medicinal plant raw material to
alcohol and water extracts\\

\begin{scriptsize}
\begin{tabular}{lccccccc}
\hline
& Specific & Specific &\multicolumn{3}{c}{$^{137}$Cs transfer} 
&\multicolumn{2}{c}{$^{137}$Cs transfer}\\
& $^{137}$Cs& $^{137}$Cs& \multicolumn{3}{c}{to tinctures,}& 
\multicolumn{2}{c}{to water, }\\
& activity & activity & \multicolumn{3}{c}{ \%}& 
\multicolumn{2}{c}{\%}\\
\cline{4-8}
\multicolumn{1}{c}{ Herbs}& in soil, &in raw &&&&&extracts at\\
&&material,&$96.5^0$&$70^0$&$40^0$&decoctions&room\\
&$ kBq\times m^{-2}$&$kBq\times kg^{-1}$&&&&&temperature\\
 \hline
&&&&&&&\\
{\it Herb of Digitalis No1} & 925 &2.66&11.6&46.6&66.2&56.4&66.2\\
{\it Herb of Digitalis No2 }& 814 &1.48&17.6&45.3&62.8&49.3&68.9\\
\hline
\multicolumn{1}{c}{Average} &&&14.6&46.0&64.5&52.9&67.6\\
\hline
{\it Leaves of Convallaria No1 }& 296 &1.49&10.7&63.1&67.1&60,0&74.0\\
{\it Leaves of Convallaria No2 }& 777 &50.83&6.3&59.7&63.9&59.1&73.0\\
\hline
\multicolumn{1}{c}{Average} &&&8.5&61.4&65.5&59.6&73.5\\
\hline
{\it Flowers of Convallaria No1 }& 740 
&11.73&4.6&58.9&66.0&60.8&69.1\\
{\it Flowers of Convallaria No2 }& 296 
&2.81&10.0&66.5&69.4&67.3&71.5\\
{\it Flowers of Convallaria No3} & 407 
&3.64&14.0&52.7&83.2&59.6&63.0\\
\hline
\multicolumn{1}{c}{Average} &&&9.5&59.4&72.9&62.6&67.8\\
\hline
\end{tabular}
\end{scriptsize}

\newpage
Table 2 \\
Solubility of some cardiac glycosides\\

\begin{scriptsize}
\begin{tabular}{llcc}
\hline
&&\multicolumn{2}{c}{Part of solvent needed to dissolve }\\
\multicolumn{1}{c}{Glycoside}&Herb&\multicolumn{2}{c}{one part of 
glycoside}\\
\cline{3-4}
&&water&methanol\\
\hline
&&&\\
Digitoxin&\it Digitalis&40 000&70\\
Convallatoxin&\it Convallaria&1000&56\\
&&&\\
\hline
\end{tabular}
\end{scriptsize}

\end{document}